\documentclass[journal=prl,showpacs,showkeys,twocolumn,nofootinbib]{revtex4-1}

\pdfpagewidth=\paperwidth
\pdfpageheight=\paperheight

\usepackage{amsmath,graphicx}
\usepackage{verbatim}
\usepackage{color}
\usepackage{amsfonts,amsbsy,amssymb}

\begin{document}

\author{Thomas W. Rosch}
\author{Paul N. Patrone}
\email{paul.patrone@nist.gov}
\affiliation{National Institute of Standards and Technology}

\date{\today}
\title{Beyond histograms: efficiently estimating radial distribution functions via spectral Monte Carlo}

\begin{abstract}
Despite more than 40 years of research in condensed-matter physics, {\it state-of-the-art} approaches for simulating the radial distribution function (RDF) $g(r)$ still rely on binning pair-separations into a histogram.  Such methods suffer from undesirable properties, including \textcolor{black}{subjectivity,} high uncertainty, and slow rates of convergence.  Moreover, such problems go undetected by the metrics often used to assess RDFs.  To address these issues, we propose (I) a spectral Monte Carlo (SMC) method that yields $g(r)$ as an analytical series expansion; and (II) a Sobolev norm that assesses the quality of RDFs by quantifying their fluctuations.  Using the latter, we show that, relative to histogram-based approaches, SMC reduces by orders of magnitude both the noise in $g(r)$ and the number of pair separations needed for acceptable convergence.  \textcolor{black}{Moreover, SMC reduces subjectivity} and yields simple, differentiable formulas for the RDF, which are useful for tasks such as coarse-grained force-field calibration via iterative Boltzmann inversion.  \looseness=-1
\end{abstract}


\maketitle

In simulations of condensed matter systems, one can barely overstate the importance of the radial distribution function (RDF) $g(r)$.   To name only a few applications, $g(r)$ is used to (i) link thermodynamic properties to microscopic details \cite{KBuff,Newman,Allen97}; (ii) compute structure factors for comparison with X-ray diffraction \cite{Yarnell77,Ashcroft76}; and more recently, (iii) calibrate interparticle forces for coarse-grained (CG) molecular dynamics (MD) \cite{Shell12,Peter09,Faller12,Muller-Plathe2002,Noid,Reith2003}.  Indeed, the RDF is such a key property that in the past few years, much work has been devoted to estimating $g(r)$ via parallel processing on GPUs \cite{Levine11}.  Given these observations, it is thus surprising that state-of-the-art techniques still construct $g(r)$ by binning simulated pair-separations into histograms, with little thought given to developing more efficient methods \cite{Allen97,Frenkel01}.  \looseness=-1

In this letter, we address this issue by proposing a spectral Monte Carlo (SMC) method for computing simulated RDFs.   The key idea behind our approach is to express $g(r)$ in an {appropriate} basis set and determine the mode coefficients via Monte Carlo estimates.  Relative to binning, we show that this approach \textcolor{black}{decreases subjectivity of the analysis, thereby reducing both the noise in $g(r)$ and the number of pair separations needed to generate useful RDFs.}  To support these claims, we also discuss how traditional $L^2$ (or sum-of-squares) metrics are insufficient for assessing convergence of $g(r)$ and propose a Sobolev norm \cite{evans2010} as an appropriate alternative.

The motivation for this work stems from the fact that $g(r)$ is increasingly being used in settings in which the details of its functional form play a critical role.  For example, scientists now routinely simulate untested materials in an effort to tailor their structural properties without the need for expensive experiments \cite{Parth,Khand}; in such applications, {\it objectively} computing RDFs is a key task.  Along related lines, structural properties are increasingly being used to calibrate coarse-grained force-fields \cite{Shell12,Peter09,Faller12,Muller-Plathe2002,Noid,Reith2003}.\footnote{In iterative Boltzmann inversion, this is achieved by updating the $i$th correction to the CG forces $F(r)$ and energies $U(r)$ via $U_{i+1} (r) = U_i (r) + k_B T \ln \left[g_i(r)/g_t (r)\right]$, $F_i(r) = -\nabla U_i(r)$, and $g_i(r) = g_i(r,\mathbb S[F_i])$,  
where $k_BT$ is the temperature, $U_0(r) = -k_B T \ln \left[g_t(r)\right]$ for a target RDF $g_t$, and $g_i(r)$ is computed from a CG MD simulation $\mathbb S$ that uses $F_i(r)$ as the CG force \cite{Shell12,Peter09,Faller12,Muller-Plathe2002,Noid,Reith2003}.}  \textcolor{black}{The success of such strategies often relies on being able to differentiate $g(r)$, which requires that simulated RDFs be accurate and relatively noise-free.}

\textcolor{black}{In this light, we therefore emphasize that histogram-based RDFs suffer from an inability to objectively control uncertainties.  This arises for several reasons.  For one, histogram bin-sizes are {\it subjective} parameters that limit the resolution of small-scale features, and often one must trade this resolution for reduced noise.  Smoothing is sometimes used as an alternative to increasing bin-sizes, but this introduces difficult-to-quantify uncertainties that depend on the choice of method.  Moreover, finite differences and/or derivatives are known to amplify noise, which renders tasks such as CG force-field calibration more difficult.  Given that (i) a corresponding experimental RDF may be unavailable for comparison, and (ii) simulation resources are often at a premium, histogram-based approaches therefore place undue burden on modelers to obtain accurate results.} 

These observations therefore motivate us to propose
\begin{align}
g(r)\approx g_M(r) = \sum_{j=0}^M a_j \phi_j(r) \label{eq:spectraldecomp}
\end{align}
where $\phi_j(r)$ are orthogonal basis functions on the domain $[0,r_c]$, $r_c$ is a cutoff radius beyond which we do not model $g(r)$, $a_j$ are coefficients to-be-determined, and $M$ is a mode cutoff.  Formally, the  $a_j$ are given by
\begin{align}
a_j \!&=\! \int_0^{r_c} \!\!\!\!\!\!{\rm d}r  \,\, \phi_j(r) g(r)    = \int_0^{r_c} \!\!\!\!\!\!{\rm d}r  \,\, \phi_j(r) \frac{N(r)}{4\pi r^2 \rho}, \label{eq:weightcomp}
\end{align}
where $\rho$ is the bulk number density and $N(r) {\rm d}r$ is the expected number of particles in a spherical shell with radius $r$, thickness ${\rm d}r$, and a particle at the origin.  In practice, Eq.~\eqref{eq:weightcomp} cannot be evaluated analytically, since $N(r)$ is unknown.  However, MD simulations yield random pair-separations distributed according to $N(r) {\rm d}r$.  Thus, we replace Eq.~\eqref{eq:weightcomp} by its Monte Carlo estimate \cite{Robert10}\looseness=-1
\begin{align}
a_j \approx \bar a_j =  \frac{\mathcal N(r_c)}{n_{\rm pairs}}\sum_{k=1}^{n_{\rm pairs}} \frac{\phi_j(r_k)}{4\pi r_k^2 \rho}, \label{eq:approxweight}
\end{align}
where $\mathcal N(r_c)$ is the expected number of particles in a sphere of radius $r_c$ (given a particle at the origin), $r_k$ is the $k$th pair separation, and $n_{\rm pairs}$ is the total number of such separations.

In order to simplify Eq.~\eqref{eq:approxweight}, note that $n_{\rm pairs} = n_c n_{\rm ppc}$, where $n_c$ is the number of MD configurations (i.e.\ timesteps or ``snapshots") used to compute $g(r)$, and $n_{\rm ppc}$ is the number of pairs-per-configuration.  The latter is well approximated by
\begin{align}
n_{\rm ppc} \approx \mathcal N(r_c) \mathcal N_{\rm tot} / 2, \label{eq:ppf}
\end{align}
when $\mathcal N_{\rm tot}$ (the number of particles per configuration) and $r_c$ are large.\footnote{This identity arises as follows.  First, the total number of pair separations is $\binom{\mathcal N_{\rm tot}}{2} \approx \mathcal N_{\rm tot}^2/2$ when $\mathcal N_{\rm tot}\to \infty$.  Only considering pairs separated by $r\le r_c$, we reduce the total number of pairs by a factor of $\mathcal N(r_c) / \mathcal N_{\rm tot}$.  We require $r_c$ to be large enough so that the relative fluctuations in $\mathcal N(r_c)$ are small.}  Substituting Eq.~\eqref{eq:ppf} into Eq.~\eqref{eq:approxweight} yields\footnote{Interestingly, related methods have been developed for density-of-state calculations under the name ``kernel polynomial method.''  See, e.g.\ Ref.~\cite{dos}.}
\begin{align}
\bar a_j =  \frac{2}{\mathcal N_{\rm tot} n_c}\sum_{k=1}^{n_{\rm pairs}} \frac{\phi_j(r_k)}{4\pi r_k^2 \rho}. \label{eq:MCweight}
\end{align}

\begin{figure}
\includegraphics[width=8.6cm]{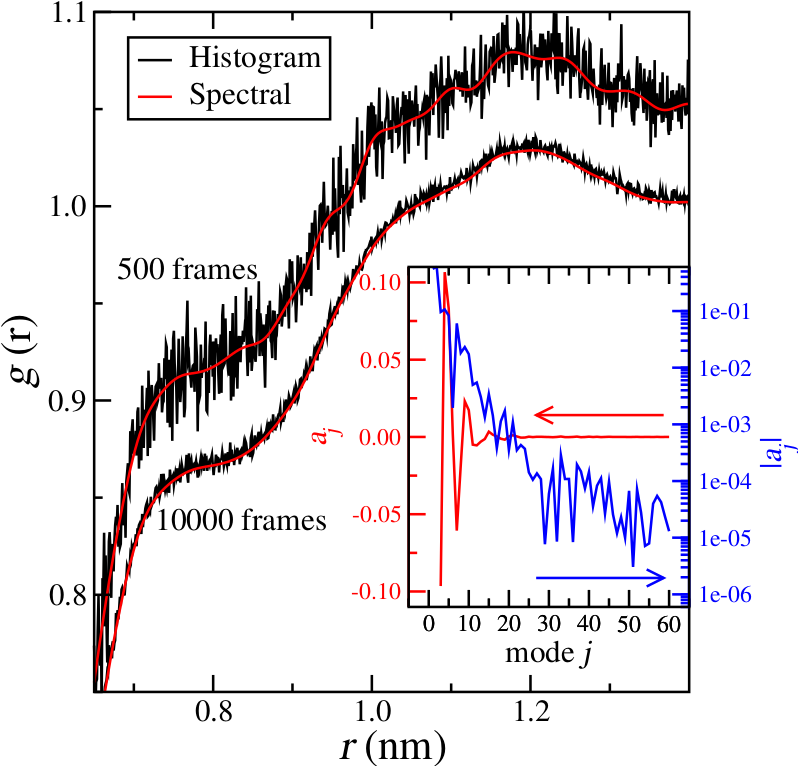}\caption{RDF of atomistic polystyrene (PS) in CG coordinates using the histogram method (black, rough curves) and SMC (red, smooth curves).  The upper and lower pairs are calculated with $n_c=500$ (shifted up by 0.05) and $n_c = 10^4$ snapshots.  The inset displays the spectral coefficients $a_j$ (left scale) and $\log|a_j|$ (right scale) for the first 60 modes.  }\label{fig:reference}
\end{figure}

We emphasize that, as opposed to histogram-based approaches,  Eq.~\eqref{eq:MCweight} provides more objective control over uncertainties in simulated RDFs.  \textcolor{red}{Specifically, for many choices of $\phi_j(r)$, the mode coefficients decay as $|a_j|< C j^{-p}$, where the constant $C$ and rate $p$ depend on the smoothness of $g(r)$.  Furthermore, for such bases, $g_M(r)$ converges to $g(r)$ uniformly in $M$} \cite{boyd2001}.\footnote{Uniform convergence of $g_M(r)$ to $g(r)$ means that for any $\epsilon$, there is an $M$ such that $|g_M(r)-g(r)|<\epsilon$ holds for all $r$.  Moreover, if $g(r)$ has $p$ derivatives, often $|a_j| \le \mathcal O(j^{-p})$; if $g(r)$ has infinitely many derivatives, the $|a_j|$ usually decay exponentially.}  This implies that in principle, the maximum error in $g_M(r)$ is controlled through $M$.  However, Monte Carlo sampling also introduces uncertainty in $a_j$, which can be estimated via
\begin{align}
\sigma_j^2 = \frac{4}{(\mathcal N_{\rm tot}n_c)^2}\sum_k \left[\bar a_j - \phi_j(r_k)/4\pi r_k^2 \rho \right]^2. \label{eq:UC}  
\end{align}
\textcolor{black}{This suggests that the largest meaningful mode cutoff $M^\star$ can be estimated from $|a_M^\star|=\mathcal O(\sigma_M^\star)$,  which corresponds to the noise-floor of $\bar a_j$; cf. Fig.~\ref{fig:reference}.  Given the uniform convergence of Eq.~\eqref{eq:spectraldecomp}, we then conclude that: (i) the error in $g_M(r)$ is the greater of either $\mathcal O(\sigma_M)$ or $\mathcal O(a_M)$ for any cutoff; and (ii) $g_M(r)$ can model all features whose characteristic size is greater than $r_c/M$.}
 
\textcolor{black}{We also emphasize that the task of choosing a suitable  basis is generally straightforward.  It is well known, for example, that if $g(r)$ is twice differentiable and $g'(0) = g'(r_c) = 0$ (which should approximately hold if $r_c$ is large enough), then $\phi_j(r) = \sqrt{2/r_c} \cos(j\pi r/r_c)$ converges uniformly and yields a series whose derivative converges to $g'(r)$ \cite{strauss1992}. Moreover, $a_j \le \mathcal O(j^{-2})$, although exponential convergence is expected when $g(r)$ is infinitely differentiable (cf. Fig.~\ref{fig:reference}) \cite{boyd2001}.  Orthogonal polynomial bases (e.g.\ Legendre or Chebyshev) are also reasonable, as they provide uniform approximations and similar rates of convergence, irrespective of boundary conditions \cite{boyd2001}.}

In order to illustrate the usefulness of Eq.~\eqref{eq:MCweight}, we compute the RDFs of a CG polystyrene (PS) model.  We first run a 10 ns, atomistic NVT simulation of amorphous, atactic PS (10 chains of 50 monomers) interacting through the pcff forcefield \cite{Sun} at 800 K and $\rho = 0.758$, with configurations output every 1 ps.  This trajectory is then mapped into CG coordinates at a resolution of 1 CG bead per monomer (located at the center of mass), so that  $\mathcal N_{\rm tot}=500$.  Next, we calculate CG RDFs via a histogram  with 1400 bins on the interval $0 \le r \le 1.4$ nm and SMC with  a cosine basis.

Figure \ref{fig:reference} shows the results of these computations for $n_c=500$ and $n_c=10^4$. The benefits of the spectral approach are readily apparent, especially when $n_c=500$.  For $n_c=10^4$, noise in the histogram method decreases by roughly a factor of $4$ or $5$ (as expected from the central limit theorem), but SMC is still dramatically smoother.  
The inset displays the first 60 spectral coefficients when $n_c=10^4$.  By eye, only 35 to 50 modes are required to reach the noise floor, far fewer degrees of freedom than the 1400 histogram bins.  

\begin{figure}
\includegraphics[width=8.6cm]{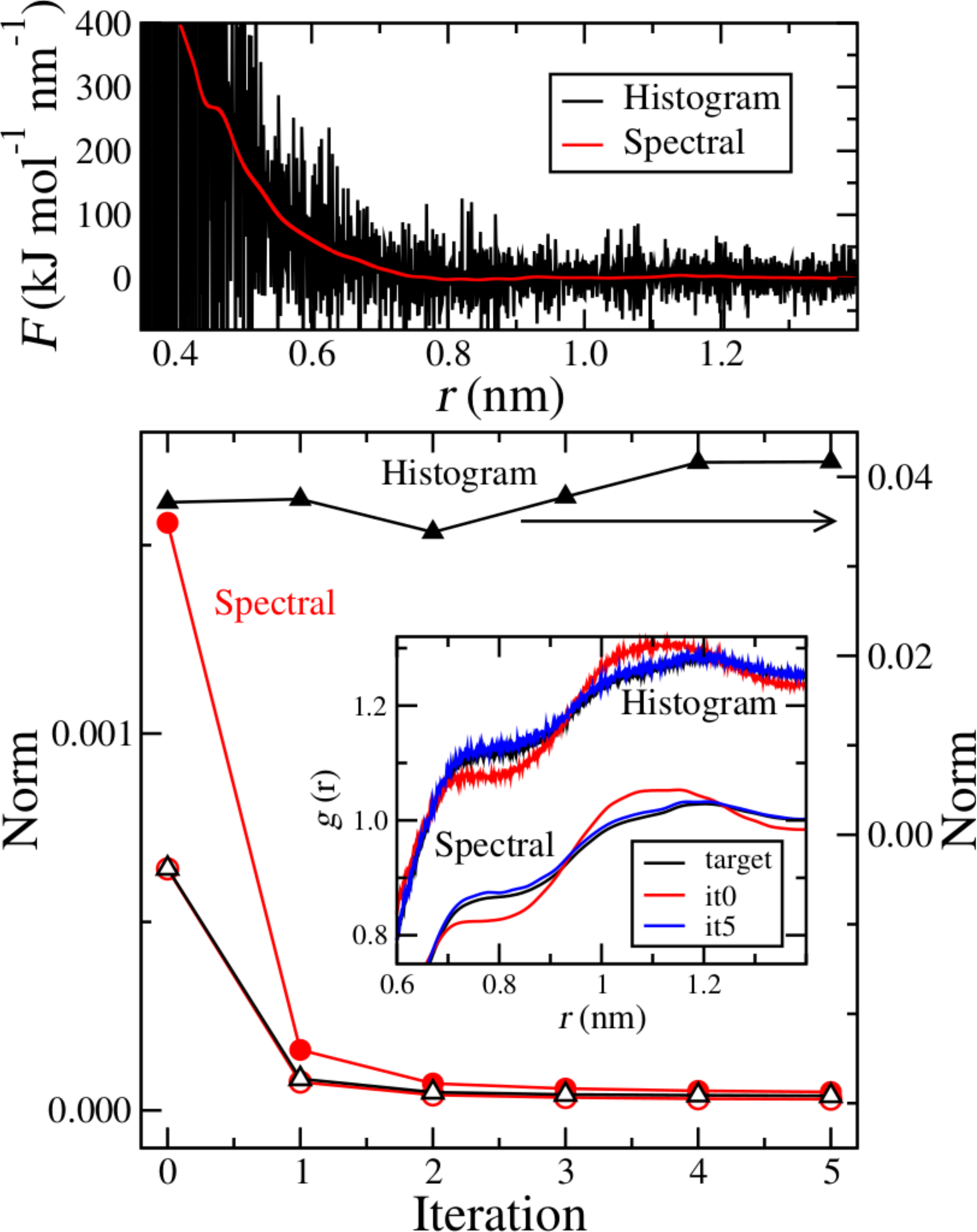}\caption{{\it Top:} CG force for PS calculated via IBI with $n_c=10^4$.  The black curve (rough) is the histogram method result, whereas the red (smooth) curve is the SMC result.  {\it Bottom:} $||g_i- g_t||_{L^2}^2$ (open symbols) and $||g_i - g_t||_{H^1}^2$ (closed symbols) norms for the histogram (circle) and SMC (triangle) methods as a function of IBI iteration.  The inset shows the corresponding RDFs. Note that $|| g -  g_t||_{H^1}^2$ for the histogram method uses the right axis and is off the scale of the left axis.}\label{fig:iterations}
\end{figure}

To further illustrate the smoothness of $g_M(r)$, we use iterative Boltzmann inversion (IBI; cf. Footnote 1) to calibrate CG forces for PS using first the histogram method and then SMC.  For the latter, we took $M=60$ and computed all forces $F_i$ analytically.  For the histogram reconstruction, we used a central finite-difference scheme to approximate the $F_i$.  IBI updates were performed using the $n_c=10^4$ RDFs in Fig.~\ref{fig:reference} as the target RDF $g_t$ (cf.\ Footnote 1).    Figure \ref{fig:iterations} shows the results of these computations.  Notably, the top subplot shows that after five iterations of IBI, the histogram-based force has extreme, high-frequency noise (despite taking $n_c=10^4$), whereas the SMC force does not. 

To make this comparison more quantitative, we define 
\begin{align}
||g||_{L^2}^2 = \frac{1}{r_c}\int_0^{r_c} \!\!\!\!\!\!{\rm d}r \,\, g(r)^2 \approx \sum_{j=1}^{n_{\rm bins}} \frac{g_j^2 \Delta r_j}{r_c}
\end{align}
where the sum is used for the histogram reconstructions, $g_j$ is the RDF evaluated in the $j$th bin, $n_{\rm bins}$ is the number of bins, and $\Delta r_j$ is the width of the $j$th bin.    Many works invoke $||g - g_t||_{L^2}$ (or variants thereof) to assess when a given RDF is sufficiently converged to $g_t$ \cite{Shell12,Iacovella2014,Faller2004}.  However, Fig.~\ref{fig:iterations} shows that both the histogram and SMC RDFs converge in $L^2$ to their respective $g_t$ at about the same rate, suggesting that this norm is not strongly affected by high-frequency fluctuations.  To account for such effects, we propose a Sobolev norm \cite{evans2010}
\begin{align}
||g||_{H^1}^2 = ||g||_{L^2}^2 + ||g'(r)||_{L^2}^2, \label{eq:h12norm}
\end{align}
where we approximate $g'(r) \approx (g_{j+1} - g_{j-1})/(r_{j+1} - r_{j-1})$ for the histogram reconstructions ($r_j$ are the bin centers).  Physically, the second term of Eq.~\eqref{eq:h12norm} assesses how smoothly $g\to g_t$.  This extra information reveals a stark difference between the histogram and SMC reconstructions insofar as the former does not improve in an $H^1$ sense.  Moreover, given that the $H^1$ and $L^2$ norms of the SMC reconstruction quickly overlap, it is clear that the difference with the $H^1$ norm of the histogram reconstruction is due to its high-frequency content.  

\begin{figure}
\includegraphics[width=8.6cm]{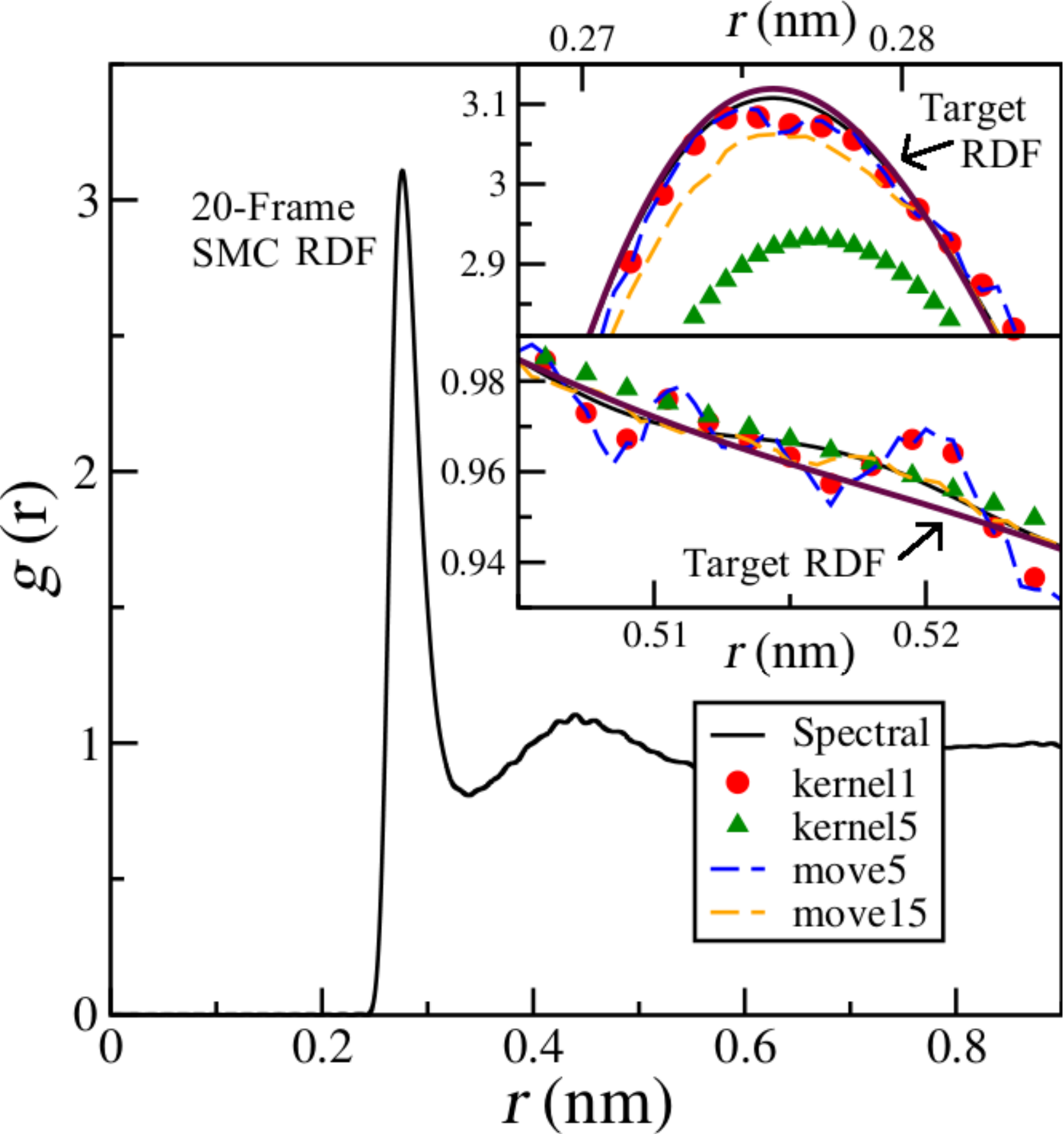}\caption{ Comparison of RDFs constructed using 20 frames of a 5000-molecule water simulation.  The main figure shows that SMC captures both the sharp peak and the rapid transition around $r=0.25$ nm.  The insets compares the 20-frame SMC, kernel-smoothed, and moving-average RDFs relative to a 200-frame SMC RDF (dark purple).  See main text for discussion. }\label{fig:water}
\end{figure}

\textcolor{black}{To test the robustness of SMC and compare with smoothing techniques, we also used 120 cosine modes to construct the O-O $g(r)$ for an TIP4P water \cite{Jorge}; cf. Fig.~\ref{fig:water}.  The atomistic system contained $\mathcal N_{\rm tot} = 5000$ molecules at 300 K and $\rho=0.8$. After 0.6 ns of equilibration, we ran a 0.2 ns production run and output configurations every 1 ps ($n_c=200$).  We take the corresponding 120-mode SMC reconstruction as a baseline for comparison, given its known convergence properties.  For histogram-based approaches, we first partitioned the domain $0\le r \le 0.9$ nm into 1800 intervals.  After binning pair-separations from the first 20 frames, we used two separate smoothing algorithms to reduce noise:\ (i) a $n$-point moving mean with $n=5$ and $n=15$; and (ii) a Gaussian-kernel that convolves the histogram with $K(x)=\exp[-0.5(x/h)^2]$ for $h=1$ pm and $h=5$ pm.  Figure \ref{fig:water} illustrates the key problem tied to the subjectivity of such methods: too little smoothing yields noisy RDFs (bottom inset), whereas too much washes out relevant features (top inset).
}

\begin{figure}
\includegraphics[width=8.6cm]{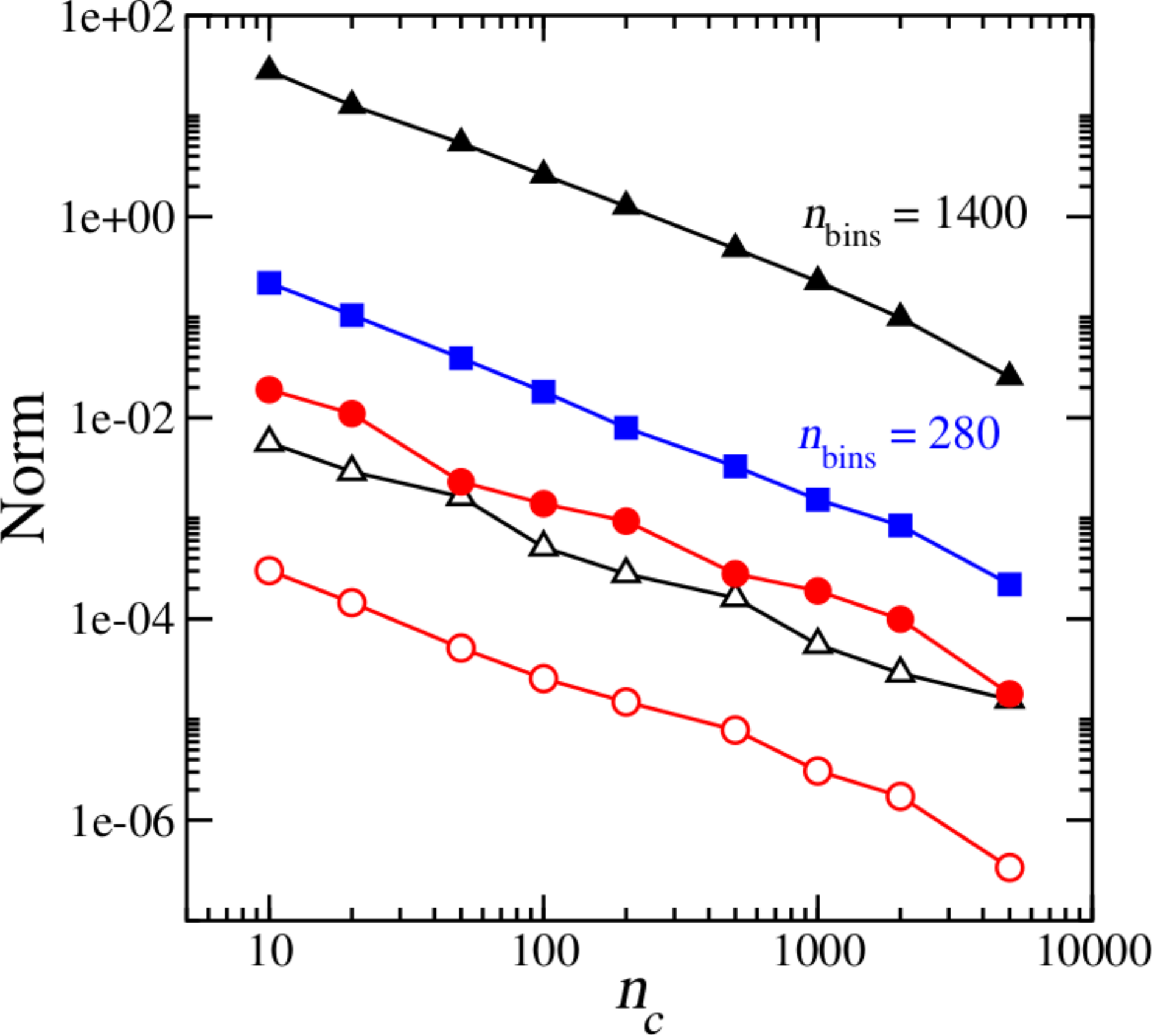}\caption{$||g - g_t||_{L^2}^2$ and $|| g -  g_t||_{H^1}^2$ as a function of $n_c$ for the 5th IBI update to the PS model in Fig.~\ref{fig:iterations}.  Here $g_t$ is the $n_c=10^4$ RDF.  Squares denote histogram estimates computed using $280$ bins.  All other symbols correspond to $1400$ bins and have the same meanings as in previous figures.}\label{fig:snapshots}
\end{figure}

This figure also suggest that as a function of $n_{c}$, SMC converges to $g(r)$ more quickly than histogram-based approaches.  To quantitatively test this, we estimated $g(r)$ for CG PS (cf.\ Figs.~\ref{fig:reference} and \ref{fig:iterations}) as a function of $n_c$ and computed the corresponding $L^2$ and and $H^1$ norms relative to the $n_c=10^4$ case (which now acts as $g_t$).  Figure \ref{fig:snapshots} shows the results of this  exercise.  Most apparent, every norm decays as roughly $1/n_c$.  Intuitively we expect this from the central limit theorem, since the variance in an average of $N$ independent, identically distributed random variables should decay as the inverse of $N$.  However, the SMC norms (circles) are at least an order of magnitude or more smaller than their histogram counterparts (squares and triangles).  This suggests that the overhead required to generate pair separations can be reduced by a factor of 10 or more simply by using SMC.  

Figure~\ref{fig:snapshots} also shows that increasing the histogram bin-width leads to {\it seemingly} smoother reconstructions of $g(r)$.  This arises from the fact that more data points contribute to any given bin, thereby decreasing fluctuations.  However, this {\it does not necessarily improve} the accuracy of such reconstructions, since  bin counts are then averages taken over increasingly large domains.\footnote{In other words, increasing the bin width trades uncertainty along the vertical axis for uncertainty along the horizontal axis.} Thus, the $H^1$ norm we propose should be used with caution, since it is likely not a valid assessment of histograms when the number of bins becomes too small.  \textcolor{black}{Along similar lines, we do not pursue quantitative comparison with convergence rates of smoothed histograms; such an analysis would require quantification of the uncertainties induced by smoothing, which can be highly non-trivial to estimate.}


Analytically, the connection between SMC and histograms can be understood by framing the latter in the context of Eq.~\eqref{eq:MCweight}.  Specifically, Eq.~\eqref{eq:MCweight} reduces to a histogram bin count when the $\phi_j(r)$ are indicator functions $I_{[r_j,r_{j+1}]}$, i.e.\ constants on an interval (i.e.\ bin) $[r_j,r_{j+1}]$ and zero otherwise.  These observations suggest that the $\phi_j$ act as a generalized histogram ``bin."  The fact that $\phi_j(r_k)$ may be non-zero for multiple $j$ indicates that each pair separation $r_k$ contributes to multiple ``bins," albeit in unequal amounts.\footnote{That is, SMC bins data according to the characteristic wavelengths with which the $r_k$ fall on the domain $[0,r_c]$.}

\textcolor{black}{From a conceptual standpoint, binning is therefore equivalent to SMC insofar as Eq.~\eqref{eq:spectraldecomp} encompasses both approaches.  Practically speaking, this suggests that both methods should be comparable in terms of computational time, which we generally find to be true.  For the trajectories analyzed in this work, single-CPU binning computations take about 20 minutes or less using custom C++ codes, whereas their SMC counterparts take about three times as long for the same value of $n_c$.  Given that MD simulations often take days, the real savings in our approach comes from needing orders of magnitude fewer pair-separations.  Moreover, SMC is embarrassingly parallel, so that the relevant computations can be reduced to a matter of minutes on standard GPUs.}

In concluding our discussion, we emphasize that despite its potential benefits, SMC nonetheless requires thoughtful implementation in order to be useful.  In particular, spectral reconstructions may give slightly negative values of $g(r)$ for small separations near $r = 0$.  This typically arises an incomplete destructive interference of the $\phi_j$ near the origin, but as Fig.~\ref{fig:water} shows, such effects are often not visually apparent.  Moreover, the problem is easily addressed by replacing Eq.~\eqref{eq:spectraldecomp} with an exponential decay $\exp(-r^k)$ near the origin (where $k$ is a fit parameter), which yields a power-law force for small separations.  We have found that this resolves any such issues with IBI when they arise.  

{\it Acknowledgments:} the authors thank Timothy Burns, Andrew Dienstfrey, and Vincent Shen for useful feedback during preparation of this manuscript.  {\it This work is a contribution of the National Institute of Standards and Technology and is
not subject to copyright in the United States.}

\bibliography{RDF}

\end{document}